\documentclass[aps,pra,showpacs,preprint]{revtex4}

\usepackage{amssymb}
\usepackage{amsmath}
\usepackage{amsfonts}
\usepackage{epsfig,subfigure,dsfont,amsthm,amsbsy,mathrsfs,amscd}
\UseRawInputEncoding
\usepackage{enumerate}
\usepackage{graphicx}
\usepackage{url}
\usepackage{bm}
\usepackage{multirow}
\usepackage{pifont}

\usepackage{float}

\newtheorem{theorem}{Theorem}

\def\QEDclosed{\mbox{\rule[0pt]{1.3ex}{1.3ex}}}

\def\QED{\QEDclosed}
\def\proof{\noindent{\bf Proof}: }
\def\endproof{\hspace*{\fill}~\QED\par\endtrivlist\unskip}
\def\la{\langle}
\def\ra{\rangle}
\def\lb{\lambda}
\def\mc{\mathcal}
\def\mr{\mathrm}
\def\be{\begin{equation}}
\def\ee{\end{equation}}
\def\ba{\begin{array}}
\def\ea{\end{array}}

\newcommand\btd{\raise 2pt \hbox{$\hat\bigtriangledown$}\hskip 1.5pt}
\newcommand\bt{\raise 2pt \hbox{$\bigtriangledown$}\hskip 1.5pt}

\begin{document}
\title{Quantum Information Masking of Hadamard Sets}

\author{Bao-Zhi Sun}
\email{sunbaozhi@qfnu.edu.cn}
\affiliation{School of Mathematical Sciences, Qufu Normal University, Shandong 273165, China}

\author{Shao-Ming Fei}
\email{feishm@cnu.edu.cn}
\affiliation{School of Mathematical Sciences,  Capital Normal University,  Beijing 100048,  China}
\affiliation{Max-Planck-Institute for Mathematics in the Sciences, Leipzig 04103, Germany}

\author{Xianqing Li-Jost}
\email{xianqing.li-jost@mis.mpg.de}
\affiliation{Max Planck Institute for Mathematics in the Sciences, Leipzig 04103, Germany}

\begin{abstract}
We study quantum information masking of arbitrary dimensional states. Given a set of fixed reducing pure states, we study the linear combinations of them, such that they all have the same marginal states with the given ones. We define the so called Hadamard set of quantum states whose Gram-Schmidt matrix can be diagonalized by Hadamard unitary matrices. We show that any Hadamard set can be deterministically masked by a unitary operation.
We analyze the states which can be masked together with the given Hadamard set using the result about the linear combinations of fixed reducing states.
Detailed examples are given to illustrate our results.
\end{abstract}


\maketitle

\section{introduction}

Due to the linearity of quantum mechanics, there are many distinguished features in quantum physics such as non-cloning \cite{1,2,3}, non-broadcasting \cite{4} and non-deleting \cite{5}. These phenomena are closely related to quantum information processing like key distribution \cite{Hwang,Scarani}, quantum teleportation \cite{Bennett,Bouwmeester} and communication security protocols \cite{Gisin,Samuel}. They are also connected to the conversation of information and the second law of thermodynamics \cite{Horodecki-FP-35-2041, Horodecki-0306044}.

Classical information encoded in composite quantum states can be completely hidden from the reduced subsystems and may be found only in the correlations. Recently, Modi et. al. investigated the same problems about quantum information \cite{Modi-PRL-120-23}. It is shown that the quantum information can be masked from the local observers. As a new kind of no-go theorems, the so-called no-masking theorem has been derived, saying that it is impossible to mask an arbitrary set of quantum states into bipartite systems such that the reduced local density matrices have no information about the quantum states.
Concerning the information masking in multipartite quantum systems, Li etc. showed that quantum states can be masked when more participants are allowed in the masking process by some schemes different from error correction codes \cite{Li}, highlighting the differences between the no-masking theorem and the classical no-go theorems \cite{Vicente}. Such theory of quantum information masking is tightly related to quantum secret sharing \cite{M,25,Zhen} and even other potential applications in quantum communication protocols.

In \cite{Li-PRA99-052343} the authors considered both deterministic and probabilistic information masking and proved that mutually orthogonal quantum states can always be served for deterministic masking. Liang et al. \cite{Liang-PRA100-030304,Liang-PRA101-042321} proved that nonzero linear operators cannot mask any nonzero measure set of qubit states and shown that the maximal maskable set of states on the Bloch sphere with respect to any maskers is the ones on a spherical circle. Moreover, Li and Modi \cite{Li-PRA102-022418} discussed the problems about probabilistic and approximate masking of quantum information. Ding and Hu discussed quantum information masking on hyperdisks and the structure of the set of maskable states \cite{DIng-PRA102-042404}. In \cite{Du-IJTP2020} the authors discussed the problem of masking quantum information encoded in pure and mixed states and found that there exists a set of four states that can not be masked, which implies that it is impossible to mask unknown pure states.

Owing to that quantum information masking has potential applications in secret sharing \cite{M,25,Zhen}, it is of importance to find out explicitly the maskable sets. Since a maskable set of quantum states may have uncountably many elements which are not orthogonal to each other, the main problem in quantum information masking is to ascertain which set of quantum states can be masked. The Hadamard matrices, consisting of unimodular entries with arbitrary phases, can be traced back to Sylvester 1867 \cite{Sylvester-PM34-1867} and Hadamard 1893 \cite{Hadamard-BCM17-1893}. Since then the construction and applications of Hadamard matrices had attracted much attention, see \cite{Tadej-OSID13-133} for review. In particular, the complex Hadamard matrices play a crucial role in the theory of quantum information \cite{Werner-JPA34-7081} such as in solving the Mean King Problem \cite{Englert-PLA284-1} and quantum tomography.

In this paper, we study the quantum information masking in terms of Hadamard matrices. We define the so-called Hadamard set of quantum states whose Gram-Schmidt matrix can be diagonalized by Hadamard unitary matrices. We prove that the Hadamard set can be masked deterministically by a unitary masker. Then we give the sufficient and necessary condition for the linear combination of the Hadamard set to be masked by the same information masker. As examples, we present the results on quantum information masking of orthonormal bases.

\section{Quantum information masking and Hadamard sets}

We denote by $\mathcal{H}_X$ the $d-$dimensional Hilbert space associated with the system $X$.
A unitary operator $\mathcal{U}$ masks the quantum
information contained in a set of states $\{|a_k\rangle_A \in \mathcal{H}_A \}_{k=1}^n$,
if it maps $|a_k\rangle_A$ to $|\Psi_k\rangle_{AB} \in \mathcal{H}_A\otimes \mathcal{H}_B, \ k=1,2,\cdots,n$
such that all the reduced states of $|\Psi_k\rangle_{AB}$ are identical,
\be\label{1}
\mathrm{Tr}_B|\Psi_k\ra_{AB}\la\Psi_k|=\rho_A,\ \ \mathrm{Tr}_A|\Psi_k\ra_{AB}\la\Psi_k|=\rho_B,\ \forall k=1,2,\cdots,n.
\ee
The reduced states
$\rho_A$ and $\rho_B$ contain no information about the value of $k$. The set $\{|a_k\rangle_A\}_{k=1}^n$ is said to be maskable with respect to the masker $\mathcal{U}$.

A set of bipartite pure states $\{|\Psi_k\ra_{AB}\in\mathcal{H}_A\otimes\mathcal{H}_B\}_{k=1}^n$ is called a set of fixed reducing states if they have identical marginal states, namely, the relations (\ref{1}) are satisfied. Given a bipartite pure state $|\Psi_0\ra_{AB}$, with Schmidt decomposition:
$$
|\Psi_0\ra_{AB}=\sum_{j=1}^r\lambda_j|\phi_j\ra_A\otimes|\psi_j\ra_B.
$$
$\sum_j\lambda_j^2|\phi_j\ra_A\la\phi_j|$ and $\sum_j\lambda_j^2|\psi_j\ra_B\la\psi_j|$ are the spectral decompositions of $\rho_A=\mathrm{Tr}_B|\Psi\ra_{AB}\la\Psi|$ and $\rho_B=\mathrm{Tr}_A|\Psi\ra_{AB}\la\Psi|$, respectively.
Conversely, if $\rho_A=\sum_j\lambda_j^2|\xi_j\ra\la\xi_j|$ and $\rho_B=\sum_j\lambda_j^2|\varsigma_j\ra\la\varsigma_j|$
are the spectral decompositions of $\rho_A$ and $\rho_B$, respectively,
the following pure bipartite state
$$
|\Psi\ra=\sum_j\lambda_j|\xi_j\ra\otimes|\varsigma_j\ra
$$
has the same reduced states as $|\Psi_0\ra_{AB}$.
If the Schmidt coefficients of $|\Psi_0\ra_{AB}$ are all not equal, then $|\phi_j\ra$ and $|\xi_j\ra$ ($|\psi_j\ra$ and $|\varsigma_j\ra$) only differ by a phase.

It has been shown in \cite{Li-PRA99-052343} that a set of fixed reduced states $\{|\Psi_k\ra_{AB}\}_{k=1}^n$ can always be written in the following form:
\be\label{FRS-schimdt}
|\Psi_k\ra_{AB}=\sum_{j=1}^r\lambda_j|\phi_j\ra_A\otimes|\psi_j^{(k)}\ra_B,~~~ k=1,2,\cdots,n,
\ee
where $\{\lambda_j\}_{j=1}^r$ are nonzero Schmidt coefficients of $|\Psi_k\ra$, $r$ is their Schmidt rank. $\rho_A=\sum_j\lambda_j^2|\phi_j\ra_A\la\phi_j|$ ($\rho_B=\rho_B^{(k)}=\sum_j\lambda_j^2|\psi_j^{(k)}\ra_B\la\psi_j^{(k)}|$) is a certain spectral decomposition of $\rho_A$ ($\rho_B$), $k=1, 2, \cdots, n$.

We first consider that, for a given set of $n$ fixed reduced states $\{|\Psi_k\ra_{AB}\}_{k=1}^n$, how to add a new state to $\{|\Psi_k\ra_{AB}\}_{k=1}^n$ so as to get a set of $n+1$ fixed reduced states.
Let
\begin{eqnarray} |\Psi(\vec{\mu})\ra&=&\sum_k\mu_k|\Psi_k\ra_{AB}=\sum_k\mu_k\sum_j\lambda_j|\phi_j\ra\otimes|\psi_j^{(k)}\ra \nonumber\\ &=&\sum_j\lambda_j|\phi_j\ra\otimes(\sum_k\mu_k|\psi_j^{(k)}\ra) =\sum_j\lambda_j|\phi_j\ra\otimes|\psi_j(\vec{\mu})\ra\label{linear-combination},
\end{eqnarray}
where
\be\label{linear-com}
|\psi_j(\vec{\mu})\ra=\sum_k\mu_k|\psi_j^{(k)}\ra,~~~j=1,2,\dots,r.
\ee
$|\psi_j(\vec{\mu})\ra$ is some linear combination of the eigenvectors of $\rho_B$ corresponding to eigenvalue $\lambda_j$. The problem is to find the conditions for
$\vec{\mu}$ such that $|\Psi(\vec{\mu})\ra$ has the same reduced states as $\{|\Psi_k\ra\}$.

\begin{theorem}\label{FRS-Lin} Let $\{|\Psi_k\ra\}_{k=1}^n$ and $|\Psi(\vec{\mu})\ra$ be the states given in \eqref{FRS-schimdt} and \eqref{linear-combination}, respectively. Then
$\{|\Psi_k\ra\}_{k=1}^n\cup\{|\Psi(\vec{\mu})\ra\}$ constitute a set of fixed reduced states if and only if
\be\label{condition}\delta_{jj'}=\la\psi_{j'}(\vec{\mu})|\psi_j(\vec{\mu})\ra
=\sum_{k,k'}\mu_k\mu_{k'}^*\la\psi_{j'}^{(k')}|\psi_j^{(k)}\ra,~~ j,j'=1,2,\cdots,r.\ee
\end{theorem}

\proof Without loss of generality, assume that
$$\omega_1=\lambda_1=\cdots=\lambda_{j_1}\neq\omega_2=\lambda_{j_1+1}=\cdots=\lambda_{j_2} \neq\omega_3\cdots\neq\omega_m=\lambda_{j_{m-1}+1}=\cdots=\lambda_r.$$
Then, for any $|\Psi(\vec{\mu})\ra$ given in \eqref{linear-combination} we have
\be\label{reduce--A}
\rho_{B}(\vec{\mu})=\mathrm{Tr}_A|\Psi(\vec{\mu})\ra\la\Psi(\vec{\mu})|
=\sum_{i=1}^m\omega_iP_i(\vec{\mu}).
\ee
For $|\Psi_k\ra$ given in \eqref{FRS-schimdt}, the reduced states have the form,
\be\label{reduce--A1}
\rho_B=\mathrm{Tr}_A|\Psi_k\ra\la\Psi_k|=\sum_{i=1}^m\omega_iP_i,
\ee
where $P_i=\sum_{j=j_{i-1}+1}^{j_i}|\psi_j\ra\la\psi_j|$ is the orthogonal projector onto the eigensubspace $\mathcal{H}_i$ corresponding to $\omega_i$,
and $P_i(\vec{\mu})=\sum_{j=j_{i-1}+1}^{j_i}|\psi_j(\vec{\mu})\ra\la\psi_j(\vec{\mu})|$ is a projector onto the same subspace as $P_i$.

From the basic acknowledge of algebra, $\rho_B=\rho_B(\vec{\mu})$ if and only if $P_i=P_i(\vec{\mu})$ for $i=1,2,\cdots,m$.
Furthermore, $\{|\psi_j\ra\}_{j=j_{i-1}+1}^{j_i}$ is an orthonormal basis for $\mathcal{H}_i$. Let $$\Psi_i=\left(\ba{ccc}|\psi_{j_{i-1}+1}\ra&\cdots&|\psi_{j_i}\ra\ea\right)$$
and
$$
\Psi_i(\vec{\mu})=\left(\ba{ccc}|\psi_{j_{i-1}+1}(\vec{\mu})\ra&\cdots&|\psi_{j_i}(\vec{\mu})\ra\ea\right),
$$
then $P_i=\Psi\Psi^\dag, \ P_i(\vec{\mu})=\Psi(\vec{\mu})\Psi^\dag(\vec{\mu})$, and we have
$$
\ba{rl}I&=\Psi^\dag\Psi=\Psi^\dag\Psi\Psi^\dag\Psi
 =\Psi^\dag\Psi(\vec{\mu})\Psi^\dag(\vec{\mu})\Psi\\
 &=\Psi^\dag(\vec{\mu})\Psi\Psi^\dag\Psi(\vec{\mu})
 =\Psi^\dag(\vec{\mu})\Psi(\vec{\mu})\Psi^\dag(\vec{\mu})\Psi(\vec{\mu})\\
 &=\left(\Psi^\dag(\vec{\mu})\Psi(\vec{\mu})\right)^2,
\ea
$$
where the fourth equality is due to that $\Psi^\dag\Psi(\vec{\mu})$ is a square matrix. Hence, $\left(\Psi^\dag\Psi(\vec{\mu})\right)^{-1}=\Psi^\dag(\vec{\mu})\Psi$. Since $\Psi^\dag(\vec{\mu})\Psi(\vec{\mu})$ is positive, we have $\Psi^\dag(\vec{\mu})\Psi(\vec{\mu})=I$.
Therefore, we obtain that $\{|\psi_j(\vec{\mu})\ra\}_{j=j_{i-1}+1}^{j_i}$ is also an orthonormal basis for $\mathcal{H}_i$, i.e., $\delta_{jj'}=\la\psi_{j'}(\vec{\mu})|\psi_j(\vec{\mu})\ra$ for $j,j'=j_{i-1}+1, \cdots, j_i$. Noting that the eigenvectors corresponding different eigenvalues are always orthogonal, we  conclude that ``only if" part is true.

Now suppose $\delta_{jj'}=\la\psi_{j'}(\vec{\mu})|\psi_j(\vec{\mu})\ra$ for $j,j'=1,2,\cdots,r$, then certainly $\rho_A(\vec{\mu})=\rho_A$. Because $\{|\psi_j\ra\}_{j=j_{i-1}+1}^{j_i}$ and $\{|\psi_j(\vec{\mu})\ra\}_{j=j_{i-1}+1}^{j_i}$ are two orthonormal bases for $\mathcal{H}_i$, it is easy to prove that $P_i=P_i(\vec{\mu})$, $i=1,2,\cdots,m$. Then we have $\rho_B(\vec{\mu})=\rho_B$, which completes the proof.
\endproof

As applications, let us consider the following two cases:

{\bf i).} $\{\lambda_j\}_{j=1}^r$ are all different. In this case, with respect to the eigenvalue $\lambda_j$, the eigenvectors $|\psi_j^{(k)}\ra$ and $|\psi_j^{(k')}\ra$ differ only by a phase.
Assume
\be\label{different-eigenvalue}
|\Psi_k\ra=\sum_{j=1}^r\lambda_j|\phi_j\ra\otimes e^{i\theta_{jk}}|\psi_j\ra,~~~ k=1,2,\cdots,n,
\ee
then $|\Psi(\vec{\mu})\ra=\sum_j\lambda_j|\phi_j\ra\otimes(\sum_k\mu_ke^{i\theta_{jk}})|\psi_j\ra$ and
$|\psi_j(\vec{\mu})\ra=(\sum_k\mu_ke^{i\theta_{jk}})|\psi_j\ra$. One has $\la\psi_{j'}^{(k')}|\psi_j^{(k)}\ra=0$ for different $j, j'$ and arbitrary $k, k'$.
The condition \eqref{condition} becomes $|\sum_k\mu_ke^{i\theta_{jk}}|=1$, $j=1,2,\cdots,n$.

{\bf ii).} $\{\lambda_j\}_{j=1}^r$ are all equal. In this case $\{|\psi_j^{(k)}\ra\}_{j=1}^r$ can be any orthonormal basis in the support of $\rho_B$ in $\mathcal{H}_B$. $|\Psi_k\ra$ can be written as,
\be\label{same-eigenvalue}
|\Psi_k\ra=\frac{1}{\sqrt{r}}\sum_{j=1}^r|\phi_j\ra\otimes |\psi_j^{(k)}\ra,~~ k=1,2,\cdots,n
\ee
One has $|\Psi(\vec{\mu})\ra=\frac{1}{\sqrt{r}}\sum_j|\phi_j\ra\otimes(\sum_k\mu_k|\psi_j^{(k)}\ra)$ and
$|\psi_j(\vec{\mu})\ra=\sum_k\mu_k|\psi_j^{(k)}\ra$. The condition that $|\Psi(\vec{\mu})\ra$ has the same reduced states as $|\Psi_k\ra$ is equivalent to that $\la\psi_{j'}(\vec{\mu})|\psi_j(\vec{\mu})\ra=\delta_{j'j}$, $\forall j', j,$ i.e., $\vec{\mu}^\dag A_{j'j}\vec{\mu}=\delta_{j'j}$, where $A_{j'j}=(\la\psi_{j'}^{(k')}|\psi_{j}^{(k)}\ra)_{k',k}$.

We now consider the quantum masking of a special set of Hadamard states.
A unitary matrix $U=(u_{jk})\in\mathcal{C}^{n\times n}$ is called Hadamard if all the entries $u_{jk}$ have the same modular $\frac{1}{\sqrt{n}}$, i.e., $u_{jk}=\frac{1}{\sqrt{n}}e^{i\theta_{jk}}$.

Given a set of states $\{|a_k\ra_A\in\mathcal{H}_A\}_{k=1}^n$, the so-called Gram-Schmidt matrix of the set is given by $G=(\la a_k|a_l\ra)_{n\times n}$.
It is well known that Gram-Schmidt matrix of a set of states is a positive Hermitian matrix. It can be diagonalized by unitary transformations. We consider special sets of states $\{|a_k\ra_A\in\mathcal{H}_A\}_{k=1}^n$.
We call $\{|a_k\ra_A\in\mathcal{H}_A\}_{k=1}^n$ a Hadamard set if the corresponding Gram-Schmidt matrix can be diagonalized by Hadamard unitary matrix, i.e., there exits a Hadamard unitary matrix $U$, such that
\be
G=(\la a_k|a_l\ra)=U^\dag\mathrm{diag}(\lambda_1,\lambda_2,\cdots,\lambda_n)U.
\ee
Obviously, any orthonormal basis of a quantum system is a Hadamard set because the related Gram-Schmidt matrix is the unit matrix.

\begin{theorem}
A Hadamard set $\{|a_k\ra_A\in\mathcal{H}_A\}_{k=1}^n$, $n\leq d$, can be deterministically masked by a unitary operation.
\end{theorem}

{\bf Proof:} Suppose $G=(\la a_k|a_{k'}\ra)=U^\dag\mathrm{diag}(\lambda_1^2, \lambda_2^2, \cdots, \lambda_n^2)U$, $U=\frac{1}{\sqrt{n}}(e^{i\theta_{jk}})$. Let
$\{|\phi_j^A\ra\}_{j=1}^n$ and $\{|\psi_j^B\ra\}_{j=1}^n$ be arbitrary orthonormal sets in $\mathcal{H}_A$ and $\mathcal{H}_B$, respectively. Set
\be|\Psi_{k}\ra=\frac{1}{\sqrt{n}}\sum_{j=1}^n\lambda_je^{i\theta_{jk}}|\phi_j^A\ra\otimes|\psi_j^B\ra, ~~ k=1,2,\cdots,n.
\ee
Then
\begin{eqnarray} \mathrm{Tr}_A|\Psi_{k}\ra\la\Psi_k|=\frac{1}{n}\sum_{j=1}^n\lambda_j^2|\psi_j^B\ra\la\psi_j^B|=\rho_B,\\
\mathrm{Tr}_B|\Psi_{k}\ra\la\Psi_k|=\frac{1}{n}\sum_{j=1}^n\lambda_j^2|\phi_j^A\ra\la\phi_j^A|=\rho_A.
\end{eqnarray}
This means that $\{|\Psi_k\ra\}$ is a fixed reduced set. Furthermore, we have:
$$
\la\Psi_k|\Psi_{k'}\ra=\frac{1}{n}\sum_j\lambda_j^2e^{i\theta_{jk'}}e^{-i\theta_{jk}}=\left(U^\dag\mathrm{diag}
(\lambda_1^2, \lambda_2^2, \cdots, \lambda_n^2)U\right)_{kk'}.
$$
Hence
$$
(\la\Psi_k|\Psi_{k'}\ra)=U^\dag\mathrm{diag}
(\lambda_1^2, \lambda_2^2, \cdots, \lambda_n^2)U=(\la a_k|a_{k'}\ra).
$$
Let $|\phi_0\ra\in\mathcal{H}_B$. Denote $|\tilde{a}_k\ra=|a_k\ra\otimes|\phi_0\ra$, $k=1,2,\cdots,n$. Then $(\la\Psi_k|\Psi_{k'}\ra)=(\la \tilde{a}_k|\tilde{a}_{k'}\ra)$.
Since two sets of states $\{|\tilde{a}_k\ra\}_{k=1}^n$ and $\{|\Psi_{k}\ra\}_{k=1}^n$ have the same Gram-Schmidt matrix, there exists a unitary operator $V$ such that $V|\tilde{a}_k\ra=|\Psi_k\ra$ for $k=1, 2, \cdots,n$. Namely, the Hadamard set $\{|a_k\ra_A\in\mathcal{H}_A\}_{k=1}^n$, $n\leq d$, can be deterministically masked.
\QED

\noindent{\bf Remark:} The Hadamard set in Theorem 2,
$$
\left\{|a_j\ra=\sum_{k=1}^d\lb_ke^{i\theta_{jk}}|\phi_k\ra,~~~|~~~j=1,2,\cdots,n\right\},
$$
in fact belongs to the set,
$$
\mc{A}=\left\{|a\ra=\sum_{k=1}^d\lb_ke^{i\theta_{k}}|\phi_k\ra|
~\mr{with~continuous~parameters}~\theta_k\in[-\pi,\pi]\right\},
$$
which was mentioned in the discussions below the theorem 4 in \cite{Modi-PRL-120-23}.
In order to investigate the structure of the set $\mc{A}$, one needs to learn the linear independent subsets of $\mc{A}$.
Certainly, the Hadamard sets present a way towards the Modi's conjecture through the study
of linear combinations of Hadamard sets.

We have shown that a Hadamard set $\{|a_k\ra\}_{k=1}^n$ can be deterministically masked by a unitary operation. Consider $|a(\vec{\mu})\ra=\sum_k\mu_k|a_k\ra$. Then $V$ transforms $|a(\vec{\mu})\ra$ to
\be
|\Psi(\vec{\mu})\ra=\sum_k\mu_k|\Psi_k\ra =\frac{1}{\sqrt{n}}\sum_j\lambda_j|\phi_j^A\ra\otimes(\sum_k\mu_ke^{i\theta_{jk}})|\psi_j^B\ra.
\ee
From Theorem \ref{FRS-Lin}, $V$ can mask the set of states $\{|a_k\ra\}_{k=1}^n$ and $|a(\vec{\mu})\ra$ together if and only if
\be\label{result-1}
\delta_{jj'}=\la\psi_{j'}(\vec{\mu})|\psi_j(\vec{\mu})\ra,\ j,j'=1,2,\cdots,n,
\ee
where $|\psi_j(\vec{\mu})\ra=\frac{1}{\sqrt{n}}(\sum_k\mu_ke^{i\theta_{jk}})|\psi_j^B\ra$. Because $\la\psi_j^B|\psi_{j'}^B\ra=\delta_{jj'}$,
\eqref{result-1} is equivalent to $\frac{1}{n}|(\sum_k\mu_ke^{i\theta_{jk}})|^2=1$. From the above analysis, we have the following result.

\begin{theorem}\label{result-masking-range} Suppose $\{|a_k\ra_A\in\mathcal{H}_A\}_{k=1}^n$ is a Hadamard set such that its Gram-Schmidt matrix is diagonalized by Hadamard unitary matrix $U$. Then $|a(\vec{\mu})\ra=\sum_k\mu_k|a_k\ra$ and $\{|a_k\ra\}_{k=1}^n$ together can be masked by some masker $V$ if and only if $ |(U\vec{\mu})_j|=1$, $j=1,2,\cdots,n$, where $\vec{\mu}=(\mu_1, \mu_2, \cdots, \mu_n)^t$.
\end{theorem}

We now present some examples to illustrate our results.

{\it Example 1. } Suppose that $\{|a_k\ra\}_{k=1}^d$ is just an orthonormal basis of $\mathcal{H}_A$.
For any Hardmard matrix $U=\frac{1}{\sqrt{d}}(e^{i\theta_{jk}})$, and orthonormal basis $\{|\phi_k\ra_A\}_{k=1}^d$, $\{|\psi_k\ra_B\}_{k=1}^d$ in $\mathcal{H}_A$ and $\mathcal{H}_B$, respectively, the fixed reduced states can be chosen to be
$$
|\Psi_k\ra_{AB}=\frac{1}{\sqrt{d}}\sum_je^{i\theta_{jk}}|\phi_j\ra_A\otimes|\psi_j\ra_B,~~ k=1,2,\cdots,d.
$$

The masker which transforms $|a_k\ra\otimes|0\ra_B$ to $|\Psi_k\ra_{AB}$, $k=1,2,\cdots,d$, can be constructed in the following way. We expand $\{|a_k\ra\otimes|0\ra_B\}_{k=1}^d$ and $\{|\Psi_k\ra_{AB}\}_{k=1}^d$ to two orthonormal bases of
$\mathcal{H}_A\otimes\mathcal{H}_B$ as $\{|a_k\ra\otimes|l-1\ra_B\}_{k,l=1}^d$ and $\{|\Psi_k\ra_{AB}\}_{k=1}^d\cup\{|\phi_k\ra_A\otimes|\psi_l\ra_B\}_{k\neq l=1}^d$, respectively. Then using $\{|a_k\ra\otimes|l-1\ra_B\}_{k,l=1}^d$ as columns, we obtain a unitary matrix $\mc{U}_1$ with the first $n$ columns given by $\{|a_k\ra\otimes|0\ra\}_{k=1}^d$. Similarly, we have $\mc{U}_2$ with the first $d$ columns given by $|\Psi_k\ra_{AB}$, $k=1,2,\cdots,d$, and the other columns given by $\{|\phi_k\ra_A\otimes|\psi_l\ra_B\}_{k\neq l=1}^d$. Then $\mc{U}_2\mc{U}_1^\dag$ is the masker which transforms $|a_k\ra\otimes|0\ra_B$ to $|\Psi_k\ra_{AB}$ for $k=1,2,\cdots,d$.

{\it Example 2. } Consider the qubit case $d=2$. Given an arbitrary linear independent set $\{|a_1\ra, |a_2\ra\}$, the GS-matrix can be written as
$$
G=\left(\begin{array}{cc}1&re^{-i\theta}\\ re^{i\theta}&1\end{array}\right).
$$
The two eigenvalues of $G$ are equal if and only if $r=0$. If $r=0$, the unitary matrix to diagonalize $G$ can be selected arbitrary. If $r\neq 0$, simple calculation shows that the elements of the eigenvectors of $G$ have the same modulus. Then $|a_1\ra, |a_2\ra$ is a Hadamard set. All the related Hadamard matrices can be written as:
$$
\frac{1}{\sqrt{2}}\left(\begin{array}{cc}e^{i\omega_1}&e^{i\omega_2}\\
e^{i(\omega_1+\theta)}&-e^{i(\omega_2+\theta)}\end{array}\right)\equiv U(\omega_1,\omega_2),~~ \forall \omega_1, \omega_2.
$$
that is, $U^\dag(\omega_1,\omega_2)\,G\,U(\omega_1,\omega_2)=\left(\ba{cc}1+r&0\\ 0&1-r\ea\right)$.
Then the corresponding fixed reduced set is of the form,
$$
\ba{l}|\Psi_1\ra=\frac{1}{\sqrt{2}}[e^{i\omega_1}|\phi_1\ra\otimes|\psi_1\ra
+e^{i(\omega_1+\theta)}|\phi_2\ra\otimes|\psi_2\ra],\\
|\Psi_2\ra=\frac{1}{\sqrt{2}}[e^{i\omega_2}|\phi_1\ra\otimes|\psi_1\ra
-e^{i(\omega_2+\theta)}|\phi_2\ra\otimes|\psi_2\ra].\ea
$$
Set $V_1=(|a_1\ra\otimes|0\ra\, |a_2\ra\otimes|0\ra)U(\omega_1,\omega_2)\left(\ba{cc}\frac{1}{\sqrt{1+r}}&0\\ 0&\frac{1}{\sqrt{1-r}}\ea\right)$. We have $V_1^\dag V_1=I_2$. Expanding $V_1$ to a unitary matrix $V=(V_1\ V_2)$ on $\mc{H}_2\otimes\mc{H}_2$, we get
\be\label{Schmidt-decomposition-1}
V^\dag(|a_1\ra\otimes|0\ra\ |a_2\ra\otimes|0\ra)U(\omega_1,\omega_2)\left(\ba{cc}\frac{1}{\sqrt{1+r}}&0\\ 0&\frac{1}{\sqrt{1-r}}\ea\right)
=\left(\ba{cc}\sqrt{1+r}&0\\ 0&\sqrt{1-r}\\ 0&0\\ 0&0\ea\right).\ee
Similarly, set $W_1=(|\Psi_1\ra\ |\Psi_2\ra)U(\omega_1,\omega_2)\left(\ba{cc}\frac{1}{\sqrt{1+r}}&0\\ 0&\frac{1}{\sqrt{1-r}}\ea\right)$. We get another unitary matrix $W=(W_1\ W_2)$,
\be\label{Schmidt-decomposition-2}
W^\dag(|\Psi_1\ra\ |\Psi_2\ra)U(\omega_1,\omega_2)\left(\ba{cc}\frac{1}{\sqrt{1+r}}&0\\ 0&\frac{1}{\sqrt{1-r}}\ea\right)
=\left(\ba{cc}\sqrt{1+r}&0\\ 0&\sqrt{1-r}\\ 0&0\\ 0&0\ea\right).\ee
From \eqref{Schmidt-decomposition-1} and \eqref{Schmidt-decomposition-2}, we have
$$V^\dag(|a_1\ra\otimes|0\ra\ |a_2\ra\otimes|0\ra)=W^\dag(|\Psi_1\ra\ |\Psi_2\ra).$$
Therefore,
$$
WV^\dag(|a_1\ra\otimes|0\ra\ |a_2\ra\otimes|0\ra)=(|\Psi_1\ra\ |\Psi_2\ra).
$$
Then, $WV^\dag$ is the corresponding masker.

Furthermore, any qubit pure state can be expressed as
$|a(u_1,u_2)\ra=u_1|a_1\ra+u_2|a_2\ra$.
From Theorem \ref{result-masking-range}, we have that the states $|a_1\ra, |a_2\ra, |a(u_1,u_2)\ra$
can be masked if and only if there exist some $\omega_i$, $i=1,2$, such that $U(\omega_1,\omega_2)\left(\begin{array}{c}u_1\\ u_2\end{array}\right)$ has unimodular elements. This is certainly true for every $|a(u_1, u_2)\ra$. Therefore, we conclude that for qubit systems, any three states can be masked by a unitary masker, which is accordance with the results in \cite{Liang-PRA100-030304}.

\section{conclusion}

The so-called no-go theorems are of great significance in information processing. No-masking is a new no-go result introduced by Modi {\it et al.} \cite{Modi-PRL-120-23}. We have studied the masking problem based on Hadamard matrices.
We have researched for which linear combinations of some fixed reducing states has the same marginal states  with the original ones. We have shown that any
set of quantum states whose Gram-Schmidt matrix can be diagonalized by Hadamard unitary matrices can be deterministically masked by a unitary operation.
The states which can be masked together with a given Hadamard set have been also investigated.
Our approach may highlight further researches on quantum information masking.

\bigskip

\noindent{\bf Acknowledgments}\, \, This work is supported by NSF of China £¨Grant No. 11701320, 12075159), Beijing Natural Science Foundation (Z190005), Academy for Multidisciplinary Studies, Capital Normal University, the Academician Innovation Platform of Hainan Province, and Shenzhen Institute for Quantum Science and Engineering, Southern University of Science and Technology (No. SIQSE202001).


\begin{thebibliography}{99}
\bibitem{1}W. K. Wootters and W. H. Zurek, {\it A single quantum cannot be cloned.} Nature (London) {\bf 299}, 802 (1982).
\bibitem{2}N. Gisin and S. Massar, {\it Optimal Quantum Cloning Machines.} Phys. Rev. Lett. {\bf 79}, 2153 (1997).
\bibitem{3}A. Lamas-Linares, C. Simon, J. C. Howell, and D. Bouwmeester, {\it Experimental Quantum Cloning of Single Photons.} Science {\bf 296}, 712 (2002).
\bibitem{4}H. Barnum, C. M. Caves, C. A. Fuchs, R. Jozsa, and B.
Schumacher, {\it Noncommuting Mixed States Cannot Be Broadcast.} Phys. Rev. Lett. {\bf 76}, 2818 (1996).
\bibitem{5}A. K. Pati and S. L. Braunstein, {\it Impossibility of deleting an unknown quantum state.}  Nature (London) {\bf 404}, 164 (2000).
\bibitem{Hwang} Won-Young Hwang, {\it Quantum Key Distribution with High Loss: Toward Global Secure Communication.} Phys. Rev. Lett. {\bf 91}, 057901 (2003).
\bibitem{Scarani} V. Scarani, H. B-Pasquinucci, Nicolas J. Cerf, M. Du\v{s}ek, N. L\"{u}tkenhaus, and M. Peev, {\it The security of practical quantum key distribution.} Rev. Mod. Phys. {\bf 81}, 1301 (2009).

\bibitem{Bennett}C. H. Bennett, G. Brassard, C. Cr\'epeau, R. Jozsa, A. Peres,
and W. K. Wootters, {\it Teleporting an unknown quantum state via dual classical and Einstein-Podolsky-Rosen channels.} Phys. Rev. Lett. {\bf 70}, 1895 (1993).
\bibitem{Bouwmeester}D. Bouwmeester, J. W. Pan, K. Mattle, M. Eibl, H. Weinfurter
and A. Zeilinger, {\it Experimental quantum teleportation.} Nature (London) {\bf 390}, 575-579(1997).
\bibitem{Gisin} N. Gisin, G. Ribordy, W. Tittel, and H. Zbinden, {\it Quantum cryptography.} Rev. Mod.
Phys. {\bf 74}, 145 (2002).
\bibitem{Samuel} Samuel L. Braunstein, and Peter van Loock, {\it Quantum information with continuous variables.}
Rev. Mod. Phys. {\bf 77}, 513 (2005)
\bibitem{Horodecki-0306044} M. Horodecki, R. Horodecki, A. Sen(De), and U. Sen, {\it No-deleting and no-cloning principles as consequences of conservation of quantum information.} arXiv: quant-ph/0306044.
 \bibitem{Horodecki-FP-35-2041} M. Horodecki, R. Horodecki, A. Sen(De), and U. Sen, {\it Common origin of no-cloning and no-deleting principles - Conservation of information.} Found. Phys. {\bf 35}, 2041 (2005).
\bibitem{Modi-PRL-120-23}K. Modi, A. K. Pati, A. Sen(De), {\it Masking Quantum Information is Impossible.} Phys. Rev. Lett. {\bf 120}, 230501 (2018).
\bibitem{Li} M. Sh. Li, and Y. L. Wang, {\it Masking quantum information in multipartite scenario.} Phys. Rev. A {\bf 98}, 062306 (2018).
\bibitem{Vicente} J. I. de Vicente, C. Spee, and B. Kraus, {\it Maximally Entangled Set of Multipartite Quantum States.} Phys. Rev. Lett. {\bf 111}, 110502(2013).


\bibitem{M} M. Hillery, V. Bu\v{z}ek, and A. Berthiaume, {\it Quantum secret sharing.} Phys. Rev. A {\bf 59}, 1829 (1999).
\bibitem{25} R. Cleve, D. Gottesman, and H. K. Lo, {\it How to Share a Quantum Secret.} Phys. Rev. Lett. {\bf 83}, 648 (1999).
\bibitem{Zhen} H. Lu, Zh. Zhang, L. K. Chen, Zh-D Li, Ch. L., Li Li, N-L Liu, X. F. Ma, Y. A. Chen, and J-W Pan, {\it Secret Sharing of a Quantum State.} Phys. Rev. Lett. {\bf 117}, 030501 (2016).


\bibitem{Li-PRA99-052343} B. Li, S. h. Jiang, X. B. Liang, X. Li-Jost, H. Fan, and S. M. Fei, {\it Deterministic versus probabilistic quantum information masking.} Phys. Rev. A {\bf 99}, 052343 (2019).
\bibitem{Liang-PRA100-030304} X. B. Liang, B. Li, and S. M.  Fei, {\it Complete characterization of qubit masking.} Phys. Rev. A {\bf 100}, 030304(R) (2019).
\bibitem{Liang-PRA101-042321} Xiao-Bin Liang, Bo Li, Shao-Ming Fei, and Heng Fan, {\it Impossibility of masking a set of quantum states of nonzero measure,} Phys. Rev. A 101, 042321 (2020).
\bibitem{Li-PRA102-022418}M. S. Li, K. Modi, {\it Probabilistic and Approximate Masking of Quantum Information.} Phys. Rev. A {\bf 102}, 022418 (2020).
\bibitem{DIng-PRA102-042404} Feng Ding, and Xueyuan Hu, {\it Masking quantum information on hyperdisks,} Phys. Rev. A 102, 042404 (2020).
\bibitem{Du-IJTP2020} Yuxing Du, Zhihua Guo, Huaixin Cao, Kanyuan Han, and Chuan Yang, {\it Masking quantum information encoded in pure and mixed states,} Int. J. Theor. Phys.(2020).
\bibitem{Sylvester-PM34-1867} J.J. Sylvester, {\it Thoughts on inverse orthogonal matrices, simultaneous sign-succesions, and tessellated pavements in two or more colors, with applications to Newton's rule, ornamental tile-work, and the theory of numbers}, Phil. Mag. {\bf 34}, 461-475(1867).
\bibitem{Hadamard-BCM17-1893} J. Hadamard, {\it Resolution d'une question relative aux determinants}, Bull Sci. Math. {\bf 17}, 240-246 (1893).
\bibitem{Tadej-OSID13-133} W. Tadej, and K. \.{Z}yczkowski, {\it A concise guide to complex Hadamard matrices}, Open Syst. Inf. Dyn. 13, 133-177 (2006).
\bibitem{Werner-JPA34-7081} R.F. Werner, {\it All teleportation and dense coding schemes}, J.Phys.A: Math. Gen. {\bf 34}, 7081-7094 (2001).
\bibitem{Englert-PLA284-1}B.G. Englert, and Y. Aharonov, {\it The mean king's problem: Prime degrees of freedom}, Phys. Lett. A {\bf 284}, 1-5(2001).
 \end{thebibliography}
\end{document}